\documentstyle[preprint,epsf,aps]{revtex}
\begin{document}

\title{The N\'eel order for a frustrated antiferromagnetic Heisenberg model: beyond
linear spin-wave theory}
\author{Qingshan Yuan$^{1,2}$}
\address{$^1$Max-Planck-Institut f\"{u}r Physik komplexer Systeme, 
N\"{o}thnitzer Str.38, 01187 Dresden, Germany\\
$^2$Pohl Institute of Solid State Physics, Tongji University, 
Shanghai 200092, P.R.China}
\maketitle
\begin{abstract}
Within Dyson-Maleev (DM) transformation and self-consistent mean-field
treatment, the N\'eel order/disorder transition is studied for an
antiferromagnetic Heisenberg model which is defined on a square lattice 
with a nearest neighbour exchange $J_1$ and a next-nearest neighbour exchange 
$J_2$ along only one of the diagonals. It is found that the N\'eel order may
exist up to $J_2/J_1=0.572$, beyond its classically stable regime. 
This result qualitatively improves that from linear spin-wave theory based on 
Holstein-Primakoff transformation.
\end{abstract}


The two-dimensional (2D) antiferromagnetic Heisenberg models have attracted 
great interest in recent years, partly because of the fact that the parent 
compounds of the high temperature superconducting materials are excellent
realizations of quasi-2D quantum antiferromagnets \cite{Chakravarty}. 
While the unfrustrated
Heisenberg model has been well understood, much attention has been paid to
the frustrated models such as square-lattice nearest and 
next-nearest neighbour interaction (so called $J_1$-$J_2$) model, triangular
lattice model and {\it Kagom\'e} lattice model etc
\cite{Oitmma,Hu,Ivanov,Mila,Singh,Lecheminant}. In these frustrated Heisenberg
models the property of the ground state, whether magnetically ordered or disordered,
is a subject of considerable interest. For example, the $J_1$-$J_2$ model takes
on N\'eel order at small $J_2/J_1$ and collinear order at large $J_2/J_1$, 
which are seperated by a region of disordered state \cite{Oitmma}. 

Very recently a $S=1/2$ Heisenberg model, which defined on a square lattice 
with a nearest neighbour antiferromagnetic exchange $J_1$ and a next-nearest 
neighbour exchange $J_2$ along only {\it one} of the diagonals of the lattice 
as shown in Fig. 1, has been proposed \cite{Zheng,Merino}. Its Hamiltonian 
is written as
\begin{equation}
H = J_1\sum_{\langle ij \rangle} {\bf S}_i\cdot  {\bf S}_j+J_2
\sum_{\langle lm \rangle} {\bf S}_l\cdot  {\bf S}_m \ \ ,\label{H}
\end{equation}
where the notation $\langle ij \rangle$ denote nearest neighbour bonds and 
$\langle lm \rangle$ denote next-nearest neighbour bonds along only one 
diagonal. Topologically this model is equivalent to the Heisenberg
antiferromagnet on an anisotropic triangular lattice \cite{Bhaumik,Trumper}.
In special cases $J_2=0,\ J_1=J_2$ and $J_1=0$, it will 
recover to unfrustrated square lattice model, 
isotropic triangular lattice model and decoupled spin chains, respectively. 
Therefore this model provides a way of
interpolating between several well-known one and two-dimensional models. One
can study the role of frustration in going from one to two dimensions.
On the other hand, this model is of direct relevance to the magnetic phases of 
some quasi-2D organic superconductors. It has been argued that this model
may describe the spin degree of freedom of the insulating phase of the layered
molecular crystals $\kappa$-(BEDT-TTF)$_2$X \cite{McKenzie}. 
The parameter $J_2/J_1$ for these materials is suggested to be $\sim 0.3-1$ 
and the magnetic frustration will play an important role. 
\begin{figure}[h]
\epsfxsize=8cm
\epsfysize=3cm
\centerline{\epsffile{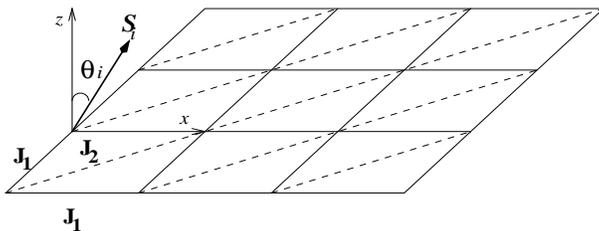}}
\caption{The 2D square lattice with nearest neighbour interaction $J_1$ and
next-nearest neighbour inteaction $J_2$ along only one of the diagonals.}
\label{fig_spin}
\end{figure}

Classically, the ground state of the model can be
derived straightforwardly as a function of the ratio $J_2/J_1$ if we assume 
that the spins lie in the $xz$ plane and are described by a spiral form 
${\bf S}_i=S(\sin \theta_i,0,\cos \theta_i)$ as shown in Fig. \ref{fig_spin}. 
Here the angle $\theta_i={\bf q}\cdot {\bf r}_i$ and the 
wavevector ${\bf q}=(q,q)$ defines a relative orientation of the spins. 
Minimization of 
the classical energy with respect to $q$ gives the result that the ground
state take on N\'eel order (i.e., $q=\pi$) for $J_2/J_1\le 1/2$, and 
spiral order with $q=arccos(-J_1/2J_2)$ for $J_2/J_1> 1/2$. 

\begin{figure}[h]
\epsfxsize=7cm
\epsfysize=4.5cm
\centerline{\epsffile{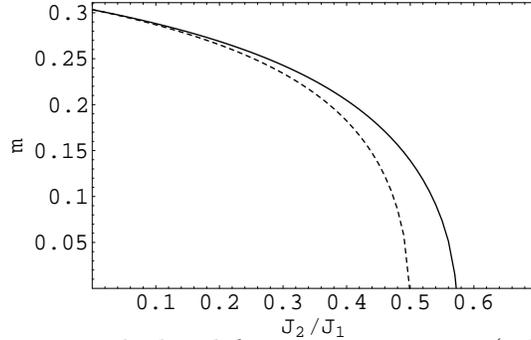}}
\caption{The magnetization $m$ calculated from our treatment (solid line) and
that from LSW theory (dashed line) as a function of $J_2/J_1$. It goes to zero
at about $J_2/J_1=0.499$ within LSW theory, but up to $J_2/J_1=0.572$ within
our treatment.}
\label{fig_Sz}
\end{figure}

The quantum model has also been studied numerically and analytically. 
The series expansions were adopted by Zheng et al. \cite{Zheng} for numerical 
calculation. It was found that the N\'eel order persists up to $J_2/J_1=0.7$. 
In the region $0.7\le J_2/J_1 \le 0.9$ there is no magnetic order and for 
larger values of $J_2/J_1$ there is incommensurate or spiral
order. It is interesting to
note that the N\'eel order exists beyond its classical result $(J_2/J_1=0.5)$.
Analytically the standard and simple linear spin-wave theory (LSW) 
based on Holstein-Primakoff (HP) transformation was used by Merino et al. 
\cite{Merino} to discuss the possible ordered and disordered states. 
It is helpful to repeat some details here for later discussion. 
First, for convenience the spin at each site is rotated along its reference
direction characterized by the angle $\theta_i$. Such a rotation may be 
accomplished by the following transformation for spin operators
\begin{eqnarray*}
S_i^x & = & \sin \theta_i \hat{S}_i^z+\cos \theta_i \hat{S}_i^x\\
S_i^y & = & \hat{S}_i^y\\
S_i^z & = & \cos \theta_i \hat{S}_i^z-\sin \theta_i \hat{S}_i^x \ 
\end{eqnarray*}
For new spins $\hat{\bf S}_i$ the reference ground state becomes 
ferromagnetic. The rotated Hamiltonian becomes
\begin{eqnarray}
H &=&J_1\sum_{\langle ij \rangle} [\cos \theta_{ij}(\hat{S}_i^x\hat{S}_j^x+
\hat{S}_i^z\hat{S}_j^z)+\sin
\theta_{ij}(\hat{S}_i^z\hat{S}_j^x-\hat{S}_i^x\hat{S}_j^z)+\hat{S}_i^y
\hat{S}_j^y]  \nonumber \\
& & +J_2\sum_{\langle lm \rangle} [\cos \theta_{lm}(\hat{S}_l^x\hat{S}_m^x
+\hat{S}_l^z\hat{S}_m^z)+\sin \theta_{lm}(\hat{S}_l^z\hat{S}_m^x-\hat{S}_l^x
\hat{S}_m^z)+\hat{S}_l^y\hat{S}_m^y] \ . 
\label{H:FM}
\end{eqnarray}
with $\theta_{ij}=\theta_i-\theta_j=q,\ \theta_{lm}=\theta_l-\theta_m=2q$. 
Then with HP transformation the above Hamiltonian with only quadratic terms 
kept may be diagonalized in the momentum space. The final dispersion relation 
for the spin excitation is 
$$\omega_{\bf k}=\sqrt{\{[J({\bf k}+{\bf q})+J({\bf k}-{\bf q})]/2-J({\bf
q})\}[J({\bf k})-J({\bf q})]}$$
with $J({\bf k})=J_1(\cos k_x+\cos k_y)+J_2\cos (k_x+k_y)$. By calculation of the
magnetization $\langle \hat{S}_i^z\rangle$ (see also the dashed line in Fig. 
\ref{fig_Sz} later) it was suggested by the authors in Ref. \cite{Merino}
that a possible N\'eel order/disorder transition happens at the point 
$J_2/J_1\simeq 0.5$. However, we want to point out the inherent limitation for
LSW theory here. From the spectrum $\omega_{\bf k}$, it is easy to find that, in
order to ensure the argument in the square root always positive in the whole 
Brillouin zone, the parameter $q$
has to be set as $\pi$ (characterizing N\'eel order) in the region
$J_2/J_1<0.5$, but must {\it not} be set as $\pi$ in the region $J_2/J_1>0.5$. 
This means that within LSW 
theory the N\'eel ordered state can never appear beyond its classically stable
region, which is in contrast with the numerical result \cite{Zheng}. 
As also pointed out by the authors in Ref. \cite{Merino} themselves, 
the interaction between spin waves becomes very
large at the transition point and it may lead to a completely different
picture for the states. So it is very necessary to go beyond the LSW theory
to see how the above result will be modified, which is the purpose of this
paper. Instead of HP transformation, the Dyson-Maleev (DM) transformation will be 
adopted to avoid $1/S$ expansion. It has been also recognized that the DM 
transformation must be prefered if one needs really to go beyond LSW theory
within a perturbation scheme \cite{Canali}. In the treatment of the so called 
$J_1$-$J_2$ model, it has been shown that the spin-wave theory based on DM 
transformation gives perfectly consistent result as that from numerical 
calculation; much better than that from LSW theory \cite{Hu}.

In this work we will not discuss the possible spiral state for large
$J_2/J_1$, but focus on the N\'eel order/disorder transition in the 
intermediate $J_2/J_1$ region, especially on the problem whether the N\'eel
order may appear beyond its classical region of stability which is one of
the most interesting topics for this model. Technically, the Hamiltonian 
under DM transformation have terms as high as sixth order when
spiral state is considered (i.e., for general $q$), which are relatively
complicated to treat. When only N\'eel state is considered the transformed
Hamiltonian has no term higher than fourth order, which can be easily treated
by mean-field (MF) theory or perturbation theory. Explicitly, one may apply
the DM transformation onto the original Hamiltonian (\ref{H}); or begin
with the rotated Hamiltonian (\ref{H:FM}) by setting $\theta_{ij}=\pi,\ 
\theta_{lm}=2\pi$ and then use the DM 
transformation for A and B two sublattices in the following form:
\begin{eqnarray*}
\hat{S}_i^{+} = (1-a_i^{\dagger}a_i)a_i,\ \  & \hat{S}_i^{-} = 
a_i^{\dagger},\ \ & \hat{S}_i^{z} = 1/2-a_i^{\dagger}a_i \\
\hat{S}_j^{-} = b_j^{\dagger}(1-b_j^{\dagger}b_j),\ \  & \hat{S}_j^{+} = 
b_j,\ \  & \hat{S}_j^{z} = 1/2-b_j^{\dagger}b_j \ , 
\end{eqnarray*}  
where $a(a^{\dagger}),\ b(b^{\dagger})$ are bosonic operators for 
sublattices A and B, respectively. Then the Hamiltonian (\ref{H:FM}) is 
transformed into
\begin{eqnarray}
H & = & -J_1\sum_{\langle ij \rangle} (1/2-a_i^{\dagger}a_i)
(1/2-b_j^{\dagger}b_j)+[(1-a_i^{\dagger}a_i)a_i b_j+
a_i^{\dagger}b_j^{\dagger}(1-b_j^{\dagger}b_j)]/2 
\nonumber \\
&+&J_2\{\sum_{\langle lm \rangle \in A} (1/2-a_l^{\dagger}a_l)
(1/2-a_m^{\dagger}a_m)+[(1-a_l^{\dagger}a_l)a_l a_m^{\dagger}+
(1-a_m^{\dagger}a_m)a_m a_l^{\dagger}]/2 + 
\sum_{\langle lm \rangle \in B} a\rightarrow b\}\ .
\label{H:DM}
\end{eqnarray}
To diagonalize the above Hamiltonian, we treat the quartic terms with a 
self-consistent MF theory. For those terms which could be decoupled in two 
ways, we will combine the two kinds of decoupling form together through
a weight factor $\lambda$ as introduced by Chu and Shen \cite{Chu}. 
For example, the term like $a_i^{\dagger}a_ib_j^{\dagger}b_j$ will be 
decoupled in the way
$$a_i^{\dagger}a_ib_j^{\dagger}b_j\simeq \lambda [\langle a_i^{\dagger}a_i
\rangle b_j^{\dagger}b_j+ a_i^{\dagger}a_i \langle b_j^{\dagger}b_j\rangle
-\langle a_i^{\dagger}a_i\rangle \langle b_j^{\dagger}b_j\rangle]+
(1-\lambda)[\langle a_i^{\dagger}b_j^{\dagger}
\rangle a_i b_j+ a_i^{\dagger}b_j^{\dagger} \langle a_i b_j\rangle
-\langle a_i^{\dagger}b_j^{\dagger}\rangle \langle a_i b_j\rangle]\ .$$
The parameter $0\le \lambda\le 1$ reflects the competition between two
decoupling ways. Its value may be decided by minimization of the energy, which was
found to be 1/2 in Ref. \cite{Chu} for unfrustrated lattice; or it is required
to be equal to 1/2 in order to keep the symmetry of the Hamiltonian before and
after decoupling \cite{Li}. We will take the value
$\lambda=1/2$ throughout our calculations. With definition of several 
parameters: $u=\langle a_i^{\dagger}a_i\rangle=\langle 
b_i^{\dagger}b_i\rangle,\ v=\langle a_i^{\dagger}b_j^{\dagger}\rangle
=\langle a_i b_j\rangle,\ w=\langle a_l^{\dagger}a_m\rangle=\langle 
b_l^{\dagger}b_m\rangle$, we may obtain a quadratic Hamiltonian in the
momentum space:
\begin{eqnarray} 
H & = & \sum_{\bf k} [C_{\bf k} (a_{\bf k}^{\dagger}a_{\bf k}+b_{\bf
k}^{\dagger}b_{\bf k})+ E_{\bf k} (a_{\bf k} b_{-\bf k}+a_{\bf k}^{\dagger}
b_{-\bf k}^{\dagger})] + {\rm const}\ ,\label{Hk}\\
C_{\bf k} & = & 2J_1(1-u+v)-J_2(1-u+w)[1-\cos (k_x+k_y)] \ ,\nonumber\\
E_{\bf k} & = & -J_1(1-u+v)(\cos k_x+\cos k_y) \ ,\nonumber\\
{\rm const} & = & {N \over 2}[J_1(-1+2u^2+2v^2-4uv)+J_2(1-2u^2-2w^2+4uw)/2]\ ,
\nonumber
\end{eqnarray}
where the summation is over half of the original Brillouin zone and 
$N$ is total number of lattice sites. Under a Bogoliubov transformation
\begin{eqnarray*}
a_{\bf k} & = & \cosh \lambda_{\bf k}\ \bar{a}_{\bf k} + \sinh \lambda_{\bf k}\
\bar{b}_{-\bf k}^{\dagger}\\
b_{-\bf k}^{\dagger} & = & \sinh \lambda_{\bf k}\ \bar{a}_{\bf k} + \cosh
\lambda_{\bf k}\ \bar{b}_{-\bf k}^{\dagger}
\end{eqnarray*}
with $\tanh 2\lambda _{\bf k}=-E_{\bf k}/C_{\bf k}$, the Hamiltonian
(\ref{Hk}) may be diagonalized into
\begin{eqnarray} 
H & = & \sum_{\bf k} [\bar{\omega}_{\bf k} (\bar{a}_{\bf k}^{\dagger}\bar{a}_{\bf
k}+\bar{b}_{\bf k}^{\dagger}\bar{b}_{\bf k}+1)-C_{\bf k}]+ {\rm const}
\end{eqnarray}
with the excitation spectrum $\bar{\omega}_{\bf k}=\sqrt{C_{\bf k}^2-E_{\bf k}^2}$. 
Correspondingly the self-consistent equations for $u,\ v,\ w$ are expressed as
\begin{eqnarray*} 
u & = & {1\over N}\sum_{\bf k}{C_{\bf k}\over\sqrt{C_{\bf k}^2-E_{\bf k}^2}} -1/2 \ ,\\
v & = & -{1\over N}\sum_{\bf k} {E_{\bf k}\cos k_x \over \sqrt{C_{\bf k}^2-E_{\bf k}^2}} \ ,\\
w & = & {1\over N} \sum_{\bf k} {C_{\bf k}\cos (k_x+k_y) \over \sqrt{C_{\bf k}^2-E_{\bf k}^2}}\ .
\end{eqnarray*}
The magnetization is simply given by $m=1/2-u$ and the ground state 
energy is $E_0=\sum_{\bf k} (\bar{\omega} _{\bf k}-C_{\bf k}) + {\rm const}$. 

\begin{figure}[h]
\epsfxsize=7cm
\epsfysize=4.5cm
\centerline{\epsffile{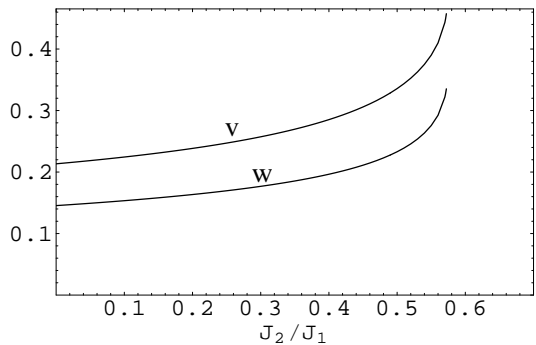}}
\caption{The self-consistent results for the parameters $v$ and $w$.}
\label{fig_vw}
\end{figure}

We show the numerical results for the above self-consistent equations in
Figs. \ref{fig_Sz} and \ref{fig_vw}. The magnetization $m$ (derived from the
parameter $u$) as a function of $J_2/J_1$
is given by the solid line in Fig. 2, which is the main result in this paper. 
It is found that the magnetization does not vanish until $J_2/J_1\simeq 0.572$, 
which gives the N\'eel order/disorder transition point. As
comparison, the result from LSW theory is plotted by the dashed line; the 
magnetization goes to zero immediately before the classical value 
$J_2/J_1=0.5$, see also Ref. \cite{Trumper}. As expected, the current result is closer
to the numerical one from series expansions and most importantly, it
qualitatively improves the result from LSW theory. As we discussed before, 
the LSW theory is impossible to deduce a N\'eel ordered
state beyond $J_2/J_1=0.5$. In our treatment the interaction between
spin waves is actually partly considered, then the fact that N\'eel order may 
exist beyond its classically stable region is shown. 
It is quite possible that the transition point
will shift to larger value if the residual interaction between spin waves is 
included. This may be also hinted from Fig. \ref{fig_vw}, where the parameter $v$,
which represents the antiferromagnetic correlation between the two original
nearest-neighbour spins, is large near the transition point. 

\begin{figure}[h]
\epsfxsize=7cm
\epsfysize=4.5cm
\centerline{\epsffile{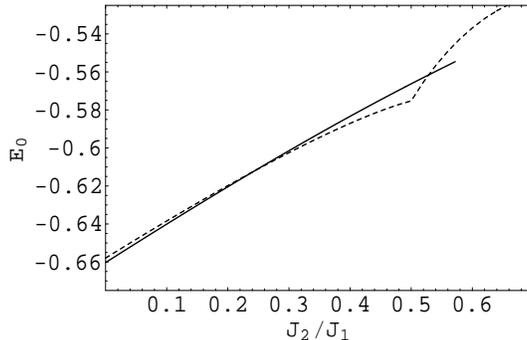}}
\caption{The ground state energy in unit of $NJ_1$ calculated from our 
treatment (solid line) and that from LSW theory (dashed line) as a function 
of $J_2/J_1$ (see text).}
\label{fig_GE}
\end{figure}

For completeness, the ground state energy in the 
region $0<J_2/J_1<0.57$ is also shown in Fig. \ref{fig_GE} by the solid line, 
which is close to the result from LSW theory (the dashed line). 
In the small $J_2/J_1$ region the energy
calculated here is a little lower than that of LSW theory, and becomes a
little higher with increase of $J_2/J_1$. For $J_2/J_1>0.5$ the energy within
LSW theory is derived from spiral state; there is a cusp at point
$J_2/J_1=0.5$ \cite{Merino}. In the current case the state is still N\'eel
ordered and the energy changes smoothly. 

In summary, within Dyson-Maleev (DM) transformation and self-consistent mean-field
treatment, we have studied an antiferromagnetic Heisenberg model on a square
lattice which includes a nearest neighbour 
exchange $J_1$ and a next-nearest neighbour exchange $J_2$ along only one of
the diagonals. This model should be of direct relevance to some layered
organic superconductors. In this work we focus on the discussion of 
N\'eel order/disorder transition for not large $J_2/J_1$. It is found that 
the N\'eel order may exist up to $J_2/J_1=0.572$, beyond its classically
stable regime. This property is consistent with numerical finding
from series expansions, which is one of the most interesting features for this
model. Especially, because the interaction between spin waves is partly 
considered in our treatment, the result derived here qualitatively
improves that from LSW theory based on Holstein-Primakoff 
transformation. It is certainly necessary to continue this work to
study the large $J_2/J_1$ region where a spiral order will appear, so that a
whole phase diagram may be constructed. 

\bigskip
The author would like to thank J. Merino for helpful discussion. This work was
supported in part by Chinese NSF. 

\end{document}